\documentclass[aps,preprint,showkeys]{revtex4-1}
\usepackage{graphicx}
\usepackage{amsmath}
\usepackage{makeidx}
\usepackage{amsfonts}
\usepackage{amssymb}
\usepackage{mathtools}
\usepackage{xcolor}

\begin{document}

\title{Distance from home matters: Investigation of a basic movement strategy}

\author{Mohsen Ghasemi Nezhadhaghighi}
\affiliation{Department of Physics, College of Science, Shiraz University, Shiraz 71454, Iran}

\author{Yahya Khalili}
\affiliation{Department of Physics, College of Science, Shiraz University, Shiraz 71454, Iran}

\author{Behafarid Hemmatpour}
\thanks{Work done while the author was affiliated with Shiraz University.}
\affiliation{IMDEA Networks Institute, Universidad Carlos III de Madrid, Madrid, Spain}

\author{Abolfazl Ramezanpour}
\email{aramezanpour@gmail.com}
\affiliation{Department of Physics, College of Science, Shiraz University, Shiraz 71454, Iran}
\affiliation{Leiden Academic Centre for Drug Research, Faculty of Mathematics and Natural Sciences, Leiden University, PO Box 9500-2300 RA Leiden, The Netherlands}

\begin{abstract}
Discovering the fundamental dynamical rules that generate the main statistical features of human mobility is essential for understanding the mechanisms underlying such processes. A prominent example is the exploration and preferential return model and its generalizations, which successfully reproduce several empirical findings. Here, we exploit another observation: the endpoint distances of a trip from the trajectory’s starting point are strongly correlated. We consider a movement process in which each user performs a sequence of trips to satisfy a set of demands, given a spatial distribution of suppliers on a two-dimensional lattice. In each trip, destinations are chosen with a probability that depends on the ratio of the initial and final distances from the user’s origin (home). We show that even a single agent with uniformly distributed demands and suppliers qualitatively reproduces  key empirical statistics, such as the power-law distribution of traveled distances. The results are also robust to introducing interactions between agents via queues and incorporating more realistic demand and supplier distributions.
\end{abstract}

\maketitle

\section{Introduction}\label{S0}
Human trajectories exhibit a broad range of spatial movements and providing an accurate statistical description of these displacements remains a challenging task. A quantitative understanding of human mobility patterns is essential due to their applications in urban planning, transportation management, and epidemiology \cite{Gonzalez2008,Batty2013,Kraemer2020}. Human mobility is described using several fundamental metrics, including  displacement and waiting time distributions, exploration rates, and visitation frequencies. Empirical observations have shown that many of these quantities follow power-law distributions, indicating the existence of universal scaling relationships underlying human mobility patterns \cite{Bettencourt2021,Gonzalez2022}. 

Remarkable advancements in location-tracking technologies using the mobile phone data, GPS navigation, and internet based platforms (e.g. Foursquare and Yelp check-ins) have enabled large scale empirical studies of mobility patterns \cite{Barbosa2018,Pappalardo2023}. These datasets provide detailed spatiotemporal dynamics of urban movements offering valuable insights into the underlying structure and dynamics of cities \cite{Barthelemy2016}. Empirical studies have revealed that trajectories of individuals within these systems follow consistent scaling patterns across diverse spatial and social contexts  \cite{Chowell2003,Brockmann2006,Alessandretti2017,Alessander2020}.

Human movement exhibits a wide range of step lengths, influenced by factors such as transportation constraints and individual social preferences \cite{Bettencourt2007}. Although human mobility shares some superficial similarities with continuous time random walk models, it significantly deviates from the assumptions of random movement due to behavioral and spatial complexities \cite{Metzler2000,Rhee2011,Zhao2015,Gallotti2016}. This has motivated studies that are based on exploration of new locations and preferential return to familiar sites in order to describe movement processes \cite{Song2010,Pappalardo2015}. Generalizations of this model have considered more details to better reflect the movement patterns and social interactions \cite{uvlaw,switching}. These studies however mostly ignore the mechanisms and primary motivations that are behind such movement processes \cite{Wu2014,Lee2015,Baumann2025,Li2025}. For instance, we know that human mobility patterns are influenced not only by external factors such as socioeconomic conditions but also by internal constraints, particularly a relatively constant energy expenditure by human body or vehicle \cite{kolbl2003, Wang2022}. Individuals tend to maintain a fixed daily energy budget for travel, leading to universal patterns in trip distribution \cite{kolbl2003, kolbl2021}.

\begin{figure}
	\includegraphics[width=16cm]{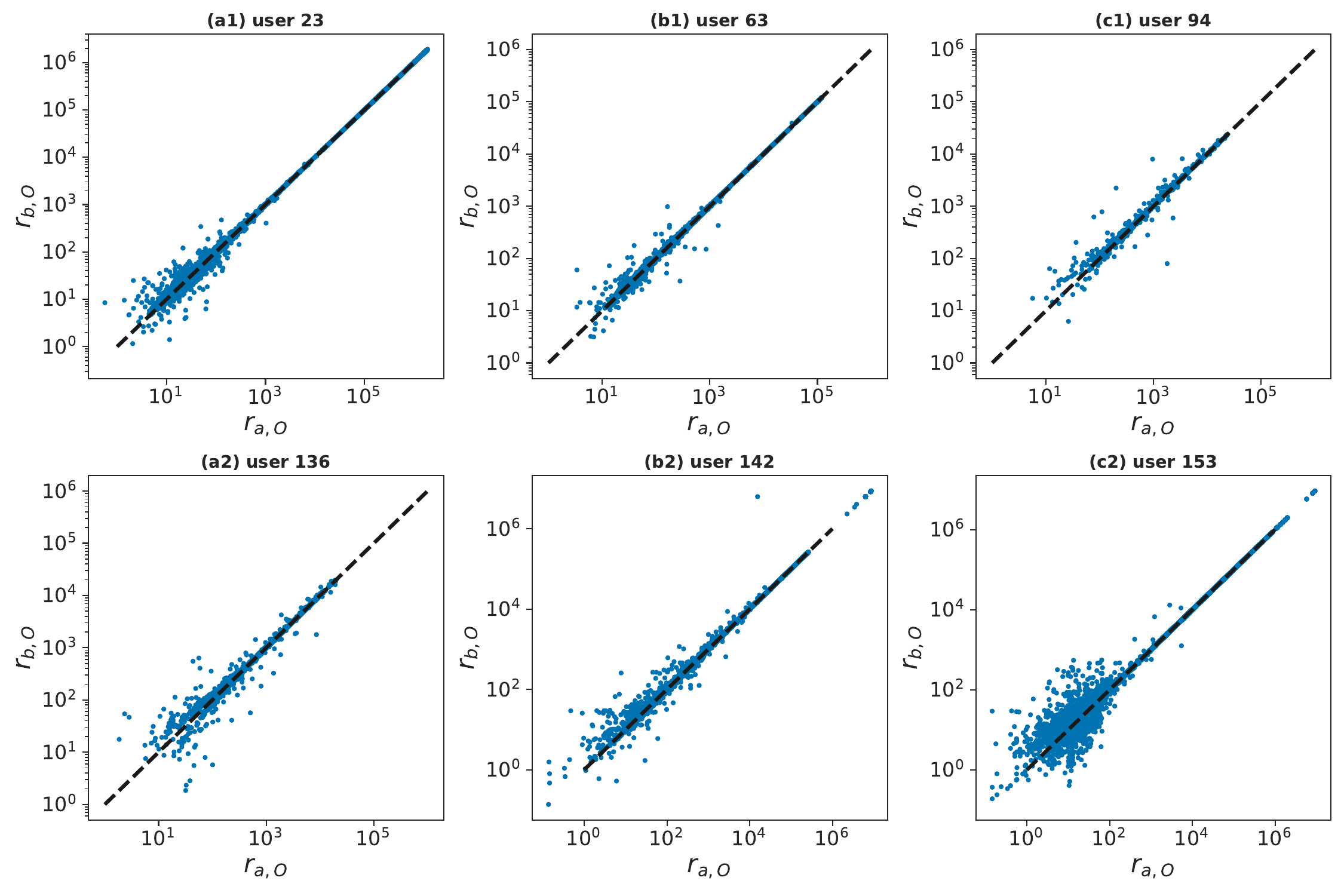} 
	\caption{Empirical data for (Euclidean) distances $r_{b,O}, r_{a,O}$ (in meters) from origin $O$ in a single trip from location $a$ to $b$. The data are from \cite{data-1,data-2,data-3}. For each user we take the trips in all the recorded trajectories for that user. The dashed lines show the linear function $r_{b,O}=r_{a,O}$. In all reported data the Pearson correlation coefficient between the two variables $(r_{a,O},r_{b,O})$ is greater than $0.99$.}\label{f0}
\end{figure}

In this paper, we introduce a minimal mobility model based on the empirical observation that the initial and final distances of a trip from the trajectory’s starting point (home) are strongly correlated (see Fig. \ref{f0}). Similar correlations have also been reported in Ref. \cite{inflation}, where the authors show that the size of modules in a trajectory network increases with the distance of the module from home position. Motivated by these findings, we propose a dynamical rule for selecting a destination in a trip, instead of imposing a power-law distribution for travel distances a priori. This study aims to demonstrate that a simple movement dynamic can reproduce several key statistical features of human mobility, including the distributions of travel distances, visitation frequencies, and exploration rates \cite{Song2010}. While we do not expect this rule to capture every aspect of human movement, it can be readily incorporated into more comprehensive models of these processes.

The paper is structured as follows: Section \ref{S1} introduces the mobility model, describing its dynamics and the main statistical quantities of interest. In Sec. \ref{S2}, we present an effective model of a single agent to focus on the sole impact of the movement strategy on the main quantities of the process. In Sec. \ref{S3}, we study the effects of interactions and distribution of suppliers in a more detailed model of interacting agents and report the results of numerical simulations. Finally, Sec. \ref{S4} summarizes the main conclusions of the study.

\section{Problem Statement}\label{S1}
In this section we provide a general statement of the problem. More details are given later where we study more specific models. The key point that we are going to exploit in this work is displayed in Fig. \ref{f0}. We use the empirical data from the trips in the GPS trajectory dataset which was collected in Geolife project \cite{data-1,data-2,data-3}. See Appendix \ref{app} for more details. Given a user trajectory, we take the initial and final distances of all trips from the starting point (or origin $O$) in that trajectory. We observe significant positive correlations between (Euclidean) distances $r_{b,O}$ and $r_{a,O}$ in a trip from position $a$ to $b$. The fact that distance of the next stop from origin in a trip is proportional to distance of the initial site is somehow expected; it is more likely to have a large distance from home if a user is already away from home. Therefore, we shall assume that the probability of going from $a$ to $b$ is a function of the ratio $r_{b,O}/r_{a,O}$.  

We consider a two-dimensional square lattice of linear size $L$, consisting of $N=L\times L$ sites. The population distribution is generated using the growth model of Ref. \cite{Li2017}. The model starts with a seed of population at the center of the lattice. At each time step, a unit of population is added to a random position of population $m_a$ with probability $\propto (m_a + C)$ only if there exists a populated site at distance less than $R_0$ from site $a$. In the following, we set $C = 1$, $R_0 = 1$. The process continues as long as the population density is less than $\rho=M/N$. At the end of this growth process, we have $m_a$ units of population (agents or users) at sites $a$ and total population $M=\sum_{a=1}^N m_a=\rho N$. This distribution defines the origin (or home) $O_i$ of the agents for $i=1,\dots,M$.

Each user $i$ has a list of demands $\mathbf{D}_i$ which is a subset of the list of possible demands denoted by $\mathbf{D}$. Each site $a$ supplies a list of demands $\mathbf{D}_a$ which is again a subset of $\mathbf{D}$, depending possibly on the population and demand distributions. The users leave their origins at time step $t=0$ to fulfill their demands. At each time step (trip) $t$, user $i$ at site $a$ selects randomly one of its demands $d\in \mathbf{D}_i$ and chooses a site $b$ which could supply that demand (i.e., $d\in \mathbf{D}_b$) with probability $\propto \exp\left(-J(d)\frac{r_{b,O_i}+1}{r_{a,O_i}+1}\right)$, see Fig. \ref{home} (panel (a)). This is the main ingredient of the models that are studied in this work. The constant $1$ in the ratio is added to avoid division by zero. We assume that the energy cost of moving from $a\to b$ is directly proportional to the distance, i.e., $\Delta E_{ab}=\Delta r_{ab}$. A demand which is supplied at site $b$ may not satisfy the user and there could be an associated waiting time $\tau_{b}$. The total travel time of an agent $t_i$ is used for the sum of waiting times in all its trips.

The question is how much of the observed behaviors of movement processes can be recovered from the above model?
The main quantities which we study in the following are time dependence of the mean-square of displacements, distribution of distances traveled in a trip, visitation frequencies, and the relation between probability of exploration and number of distinct visited sites \cite{Song2010}.

\section{An effective model}\label{S2}
In this section, we consider a single user to focus mainly on the impact of the movement strategy on the statistical behaviors of the process. 

\begin{figure}
	\includegraphics[width=12cm]{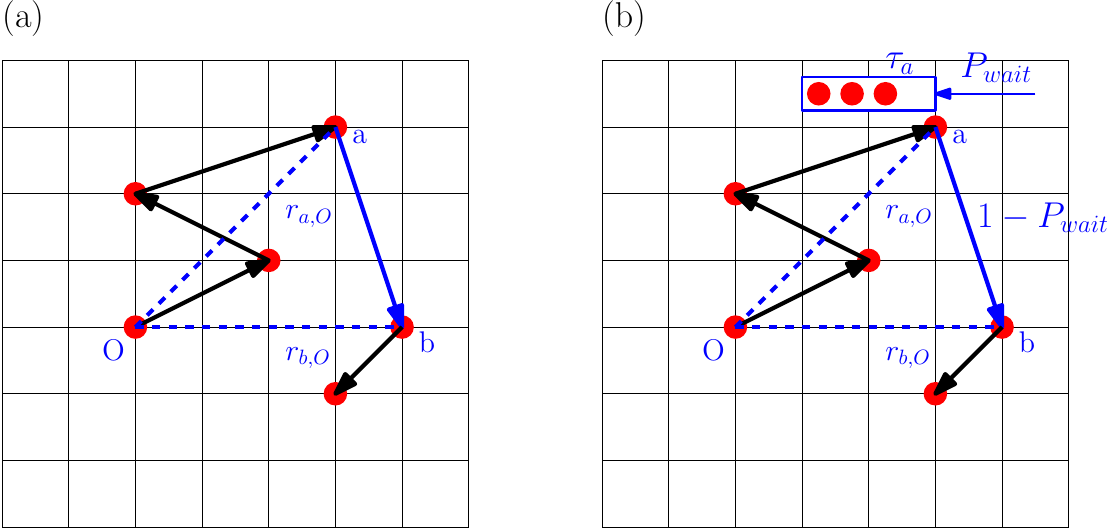} 
	\caption{Illustration of the movement strategy. The probability of going from $a$ to $b$ in a trip $P_{a\to b}\propto \exp(-J\frac{r_{b,O}+1}{r_{a,O}+1})$ depends on the ratio of distances $r_{b,O}, r_{a,O}$ from origin $O$ (home). Panel (a): a single agent moves according the above rule with no waiting time. Panel (b): the agents, depending on their waiting time $\tau_a$ in the queue, either remain in the queue with probability $P_{wait}=e^{-\gamma \tau_a/\tau_0}$ to satisfy a given demand or move to another location for this purpose.}\label{home}
\end{figure}

The agent starts its movement from the origin $O$ of a two-dimensional square lattice of size $N=L\times L$. At each step (trip) $t$ the user selects a demand $d$ with probability $P_t(d)$ from a list of $D$ distinct demands. Here we assume that all sites are able to supply any demand $d$. The probability of going from site $a$ to $b$ is
\begin{align}
	W_t( a \to b) \propto \sum_d P_t(d)\exp{[-J(d)\left(\frac{r_{b,O}+1}{r_{a,O}+1}\right)]},
\end{align}
where $O$ is the starting point (origin) and $J(d)$ controls the closeness of destination $b$ to the origin. Intuitively, small $J$ makes all destinations almost equally likely (the agent ventures far from home), while
a large $J$ favors destinations closer to home than the current location. An illustration of the movement is presented in Fig. \ref{home} (panel (a)). All distances here are Manhattan distances in the two dimensional lattice. In the following we simply take $P_t(d)=1/D$ that is the demands are chosen randomly and uniformly. The probability of finding the user at site $a$ evolves with time step $t$ as follows
\begin{align}
	P_{t+1}(a) = \sum_{b=1}^N W_t( b \to a) P_t(b).
\end{align}
In this way, the probability distribution of the distance $\Delta r$ traveled in step $t$ is given by
\begin{align}\label{dr}
	P_t(\Delta r) = \sum_{a}\sum_{b}P_{t}(a) W_t(a\to b)\delta_{\Delta r,\Delta r_{ab}}.
\end{align}

\begin{figure}
	\includegraphics[width=16cm]{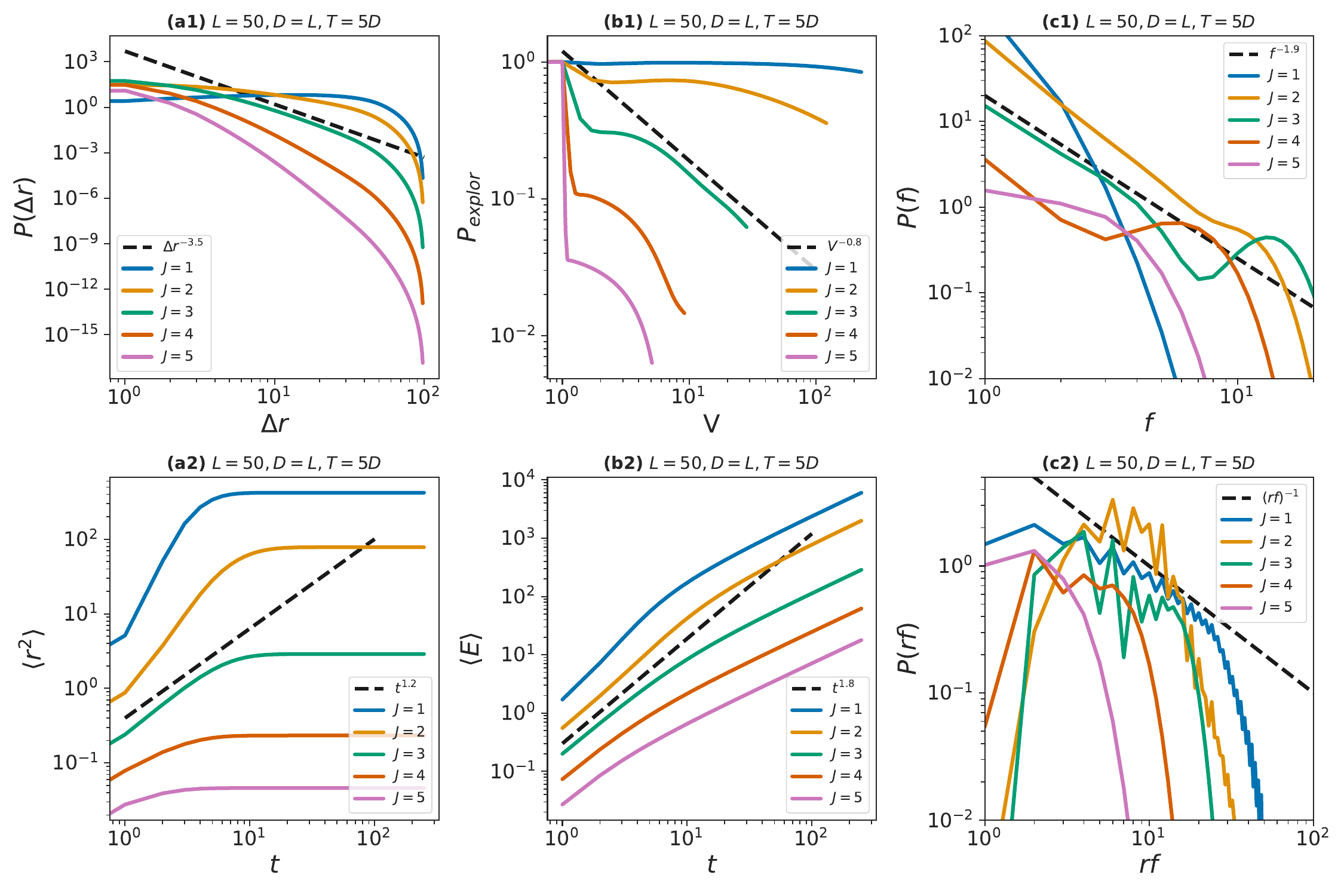} 
	\caption{Main quantities of the effective model for $T=5L$ time steps with $L=50, D=L$. Panel (a1): probability distribution of trip distances $\Delta r$. Panel (b1): probability of exploration vs the number of distinct visited sites $V$. Panel (c1): probability distribution of visit frequencies $f$. Panel (a2): mean-square displacement vs time step $t$. Panel (b2): consumed energy vs time step $t$. Panel (c2): number of sites visited $f$ times from distance $r$ vs the product $rf$. The results are obtained by numerical solution of the equations given in Sec. \ref{S2}. The dashed lines are shown for reference alongside the results.}\label{f1}
\end{figure}

We solve the above equations numerically in a lattice of linear size $L=50$ for $T=5L$ time steps with the number of demands $D=L$.
The qualitative behaviors of the results are not very sensitive to values of $L, D$ and $T$, for sufficiently large values of these quantities. The form of function $J(d)$ is however relevant here and can change the main statistical features of the process. For simplicity, we take $J(d)=J$ that is independent of the chosen demand.

Figure \ref{f1} (panel (a1)) displays $P(\Delta r)=\sum_{t=1}^T P_t(\Delta r)$ which is obtained from numerical solution of the above equations. We see that by varying the magnitude of $J$, $P(\Delta r)$ displays very different behaviors ranging from nearly Poisson for small $J<1$, to power-law for $J\simeq 3$. The empirical data suggest a power-law scaling $P(\Delta r) \propto \Delta r ^{-1-\alpha_r}$ of exponent $0<\alpha_r\le 2$ \cite{Song2010}. This occurs for larger values of $J$, where the agent tends to exhibit a sub-diffusion movement. As Fig. \ref{f1} (panel (a2)) shows the mean-square displacement $\langle r^2\rangle$ initially grows as $t^{\nu}$ with exponent $\nu<1$ for $J>3$. In fact, for very small $J$, all destination sites $b$ have nearly the same probability which results in super-diffusion whereas for very large $J$, destinations that have smaller distances $r_{b,O}$ are dominated and we observe a sub-diffusion. Real data however suggest an ultra-slow diffusion where the mean-squared displacement scales with the logarithm of time steps \cite{Song2010}.

Next we write a recursive relation for the probability $P_{t,a}(f_a)$ that site $a$ has been visited $f_a$ times at time step $t$:
\begin{align}\label{pf}
	P_{t+1,a}(f_a) = P_{t,a}(f_a) (1-P_{t+1} (a)) + \theta(f_a) P_{t,a} (f_a-1) P_{t+1}(a).
\end{align}
$\theta(x)$ is the step function which equals $1$ for $x>0$, otherwise it is zero. Given all the probabilities we've established, we can now obtain the probability of exploration,
\begin{align}\label{exp}
	P_t^{explor} &= \sum_a P_t(a)P_{t-1,a}(f_a=0),\\
	V_t &= \sum_a (1-P_{t,a} (f_a=0)),       
\end{align}
in which $f_a =0$ means that the number of times that site $a$ is visited must be zero to have an exploration. $V_t$ is the average number of distinct sites visited by the user up to time step $t$. Figure \ref{f1} (panel (b1)) shows how exploration probability $P^{explor}$ decays with the number of visited sites $V$. For large $V$ we observe a power-law tail $P^{explor} \propto V^{-\alpha_e}$ with exponents $\alpha_e\simeq 0.8$ around $J\simeq 3$. Smaller exponents are observed for smaller values of $J$. Again, empirical data suggest $\alpha_e \simeq 0.2$ \cite{Song2010}.

From the probability $P_{t,a}(f)$ we can obtain the average number of sites with frequency of visits $f$ at time step $t=T$, that is $P(f)=\sum_a P_{T,a}(f)$. From real data we expect a power-law $P(f) \propto f^{-(1+1/\alpha_f)}$ of exponent $\alpha_f \simeq 1.2$ \cite{Song2010}. This function is reported in Fig. \ref{f1} (panel (c1)). For $J=3$ we get $(1+1/\alpha_f) \simeq 1.9$. We also look at the average number of sites that are at distance $r$ from the origin and are visited $f$ times,
\begin{align}
	P(r,f)=\sum_a \delta_{r,r_{a,O}} P_{T,a}(f)/r.
\end{align}
This is computed at time step $t=T$ and is scaled with $r$. The universal visitation law of Ref. \cite{uvlaw} states that $P(r,f)\propto (rf)^{-\alpha_u}$ with exponent $\alpha_u\simeq 2$. Figure \ref{f1} (panel (c2)) displays the sum of probabilities $P(r,f)$ for a given value of $rf$. Here for $J\simeq 1$ we observe a nearly power-law relation with exponent $\alpha_u\simeq 1$.

\begin{figure}
	\includegraphics[width=12cm]{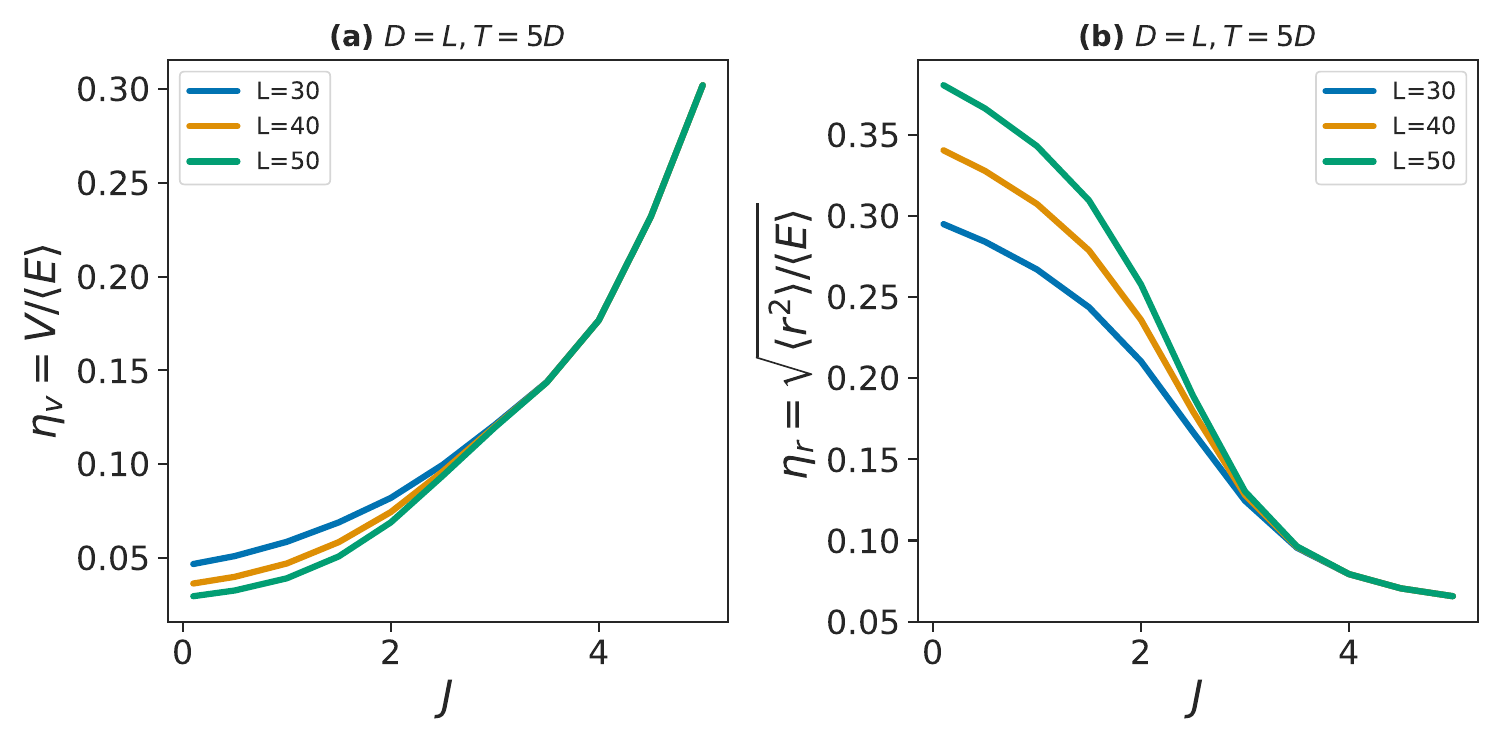} 
	\caption{Efficiencies of the effective model for $T=5L$ time steps with $L=30,40,50, D=L$. The average number of visited sites $V$ and mean-squared displacement $\langle r^2\rangle$ are compared with the average consumed energy $\langle E\rangle$. Optimal values of $J$ occur where both efficiencies are considerable. The results are obtained by numerical solution of the equations given in Sec. \ref{S2}.}\label{f2}
\end{figure}

Finally we consider the consumed energy $\langle E\rangle$ and compare it with the number of visited sites $V$ and mean-squared displacement $\langle r^2\rangle$ to characterize efficiency of the process \cite{Biazzo2020,Biazzo2021}. The consumed energy, which here is sum of the trip distances $\Delta r_{ab}$, grows quickly as $t^2$ at the beginning where still larger trip distances appear in the movement process. In the stationary state, the trip distances are of the same order and the energy increases nearly linearly with $t$. Let us define the two measures of efficiency as
\begin{align}
	\eta_{v} &=\frac{V}{\langle E\rangle},\\
	\eta_{r} &=\sqrt{\frac{\langle r^2\rangle}{\langle E\rangle}}.
\end{align}
For a given value of consumed energy, higher values of these quantities correspond to a larger number of visited sites and larger distances traveled by the agent. Figure \ref{f2} shows that $\eta_v$ increases and $\eta_r$ decreases with $J$, as expected.
The behaviors change around $J=2$ suggesting an optimal interval for the parameter $J\in (1,4)$, where both the measures have considerable values. This is the same region of the parameter that we observe power-law behaviors for the main distributions reported in Fig. \ref{f1}.

\section{Numerical Simulations: Interacting agents}\label{S3}
As shown in the previous section, although interactions among agents are neglected in the effective model, power-law behavior is still observed in different properties of the system, including the distribution functions. In this section, we investigate in greater detail the dynamics of demand satisfaction and examine how the correlation between the distance $r_{b,O}$ and $r_{a,O}$ affects the outcomes in a set of agents that interact through queue formation.

We consider an urban dynamics where each resident leaves home with a set of tasks to complete and therefore must visit various locations throughout the city to satisfy their daily needs. Service providers comprising a broad range of shops, stores, and public institutions are dispersed across the urban area, and individuals may interact with several of them over the course of a day. At each of these locations, a visitor may spend a non-negligible dwell time, which depends on multiple factors, including the number of individuals simultaneously seeking service at that facility. Reference \cite{Andreotti2025} has shown that service providers themselves exhibit a nontrivial spatial distribution across the urban environment. Using Foursquare data from the city of Bologna, they show that the probability distribution of Points of Interest (POIs) across cities exhibit clear power-law scaling when analyzed at city scale. As previously mentioned, the distribution of population in cities also follows a power-law distribution. Therefore, the distribution of urban service providers can also be modeled using the framework presented in Sec. \ref{S1}. 

There are some distinct categories for POIs across the city i.e. Business and Professional
Services, Community and Government, Dining and Drinking, Health and Medicine, Retail, and Travel and Transportation. Here we assume there are $D$ categories of POIs in the city, where ${n_1,n_2,\dots,n_D}$ are the number of each service provider that have been distributed across the city. In this simulation, as mentioned earlier, we used the model of Ref. \cite{Li2017} to generate a random and spatially heterogeneous distribution for each class of service providers across the city. Specifically, the model parameters were set to $C=1$ and $R_0=3$. Note that increasing the value of $R_0$ leads to a broader spatial spread of the distribution of service providers throughout the urban area.

Here we consider again a two-dimensional square lattice of linear size $L=70$, with the population distribution described in Sec. \ref{S1}. The population density is fixed to $\rho=M/N = 10$. Each agent starts from its home and moves according to the dynamical rule described in Sec. \ref{S1}, in order to satisfy some demands which are selected randomly and uniformly from the set of $D$ categories of POIs.

When an agent randomly samples a location $r_a$ capable of fulfilling demand type $d_a$, the site may already host a queue of length $\tau_a$. This queue emerges due to the structural imbalance between the number of demand-satisfying sites $N_D=\sum_{j=1}^D n_j$ and the significantly larger population size $M$, with $M\gg N_D$. In our simulations, the total number of service providers in the city is given by $N_D=L^2$. The number of service providers in each category is $n_j = N_D/D$, where $D$ denotes the number of distinct service categories. After observing $\tau_a$, the agent proceeds according to a probabilistic decision rule, see Fig. \ref{home} (panel (b)). The probability of remaining in the queue is prescribed by
\begin{eqnarray}\label{waitingprob}
P_{wait} = e^{-\gamma \frac{\tau_a}{\tau_0}},
\end{eqnarray}
where $\tau_0=M/N_D$. The parameter $\gamma$ sets how strongly a long queue discourages waiting, and $\tau_0$ is a reference queue length. With the complementary probability $1-P_{wait}$, the agent selects an alternative site that satisfies the same demand and relocates according to the transition probability $\propto \exp\left(-J\frac{r_{b,O}+1}{r_{a,O}+1}\right)$. Note that when $\gamma=0$, the queue length does not affect the agent’s decision; the agent remains in the queue until it is served and its demand is satisfied.

\begin{figure}
	\includegraphics[width=16cm]{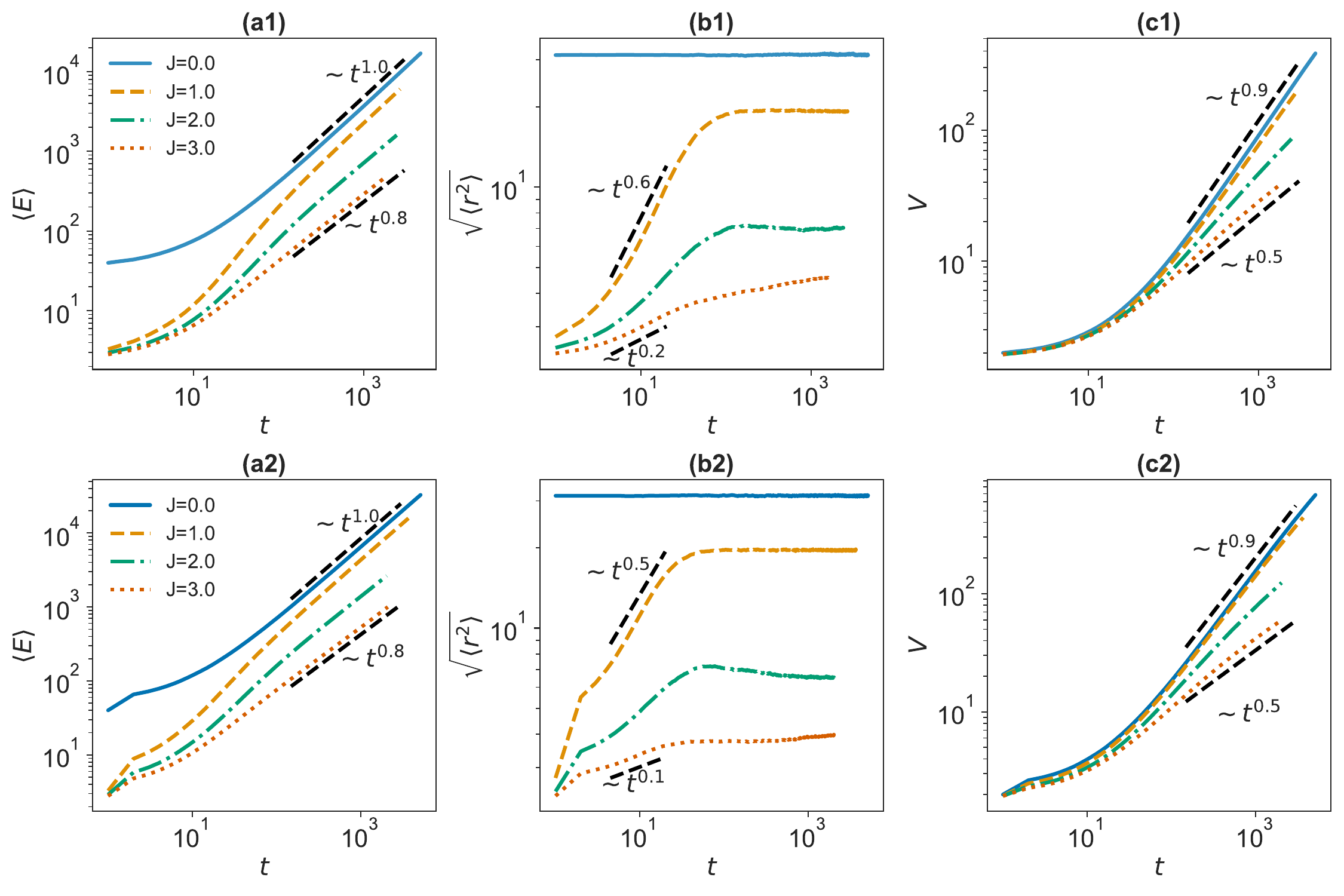} 
	\caption{ The mean consumed energy $\langle E \rangle$, the mean-squared displacement $\sqrt{\langle r^2 \rangle}$ and average number of visited sites $V$ versus time $t$ after $T=10L$ number of steps with $L=70$ and $D=20$. Top panels: Results for the case in which agents remain in the queue to satisfy a given demand, regardless of their waiting time. Bottom panels: Results for the case in which agents, depending on their waiting time in the queue, either remain in the queue with probability $P_{wait}$ to satisfy a given demand or move to another location for this purpose. The reference lines with the mentioned exponents serve as a guide to the eye.}\label{f3}
\end{figure}

\begin{figure}
	\includegraphics[width=16cm]{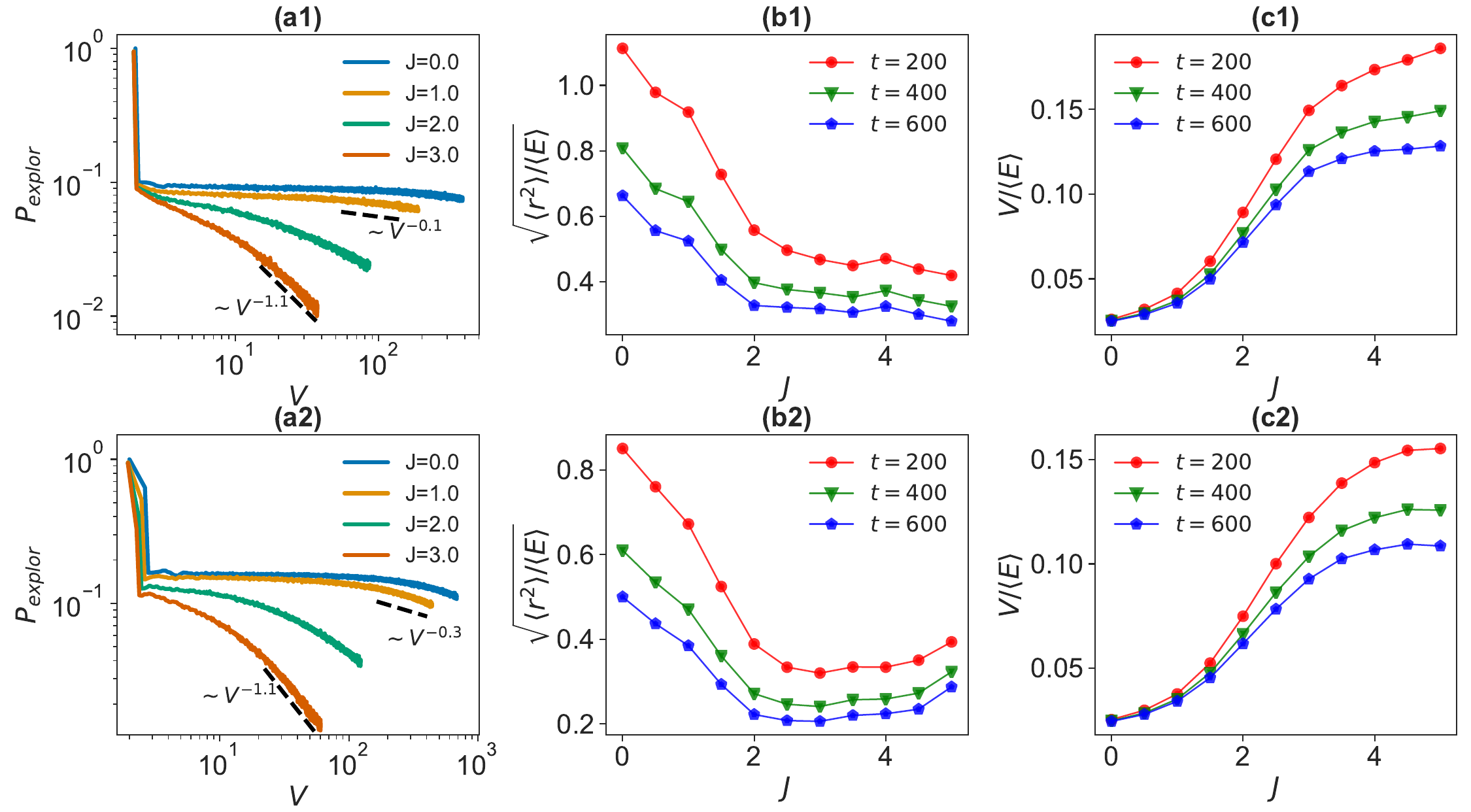} 
	\caption{Exploration probability $P_{explor}$ as a function of $ V$ and different measures of efficiencies $\eta_r = \sqrt{\langle r^2\rangle /\langle E\rangle}$ and $\eta_v = V/\langle E\rangle$ as a function of control parameter $J$. Optimal values of $J$ occur where both efficiencies are considerable. All distribution functions are obtained after $T=10L$ number of steps with $L=70$ and $D=20$. Top panels: Results for the case in which agents remain in the queue to satisfy a given demand, regardless of their waiting time. Bottom panels: Results for the case in which agents, depending on their waiting time in the queue, either remain in the queue with probability $P_{wait}$ to satisfy a given demand or move to another location for this purpose. The reference lines with the mentioned exponents serve as a guide to the eye.}\label{f4}
\end{figure}

These coupled mechanisms produce a mobility process in which both spatial and temporal components jointly determine the user’s trajectory. Such a formulation is consistent with empirical characteristics of intra-urban movements, where the perceived cost of travel reflects both physical distance and queue-induced delays. The user’s decision-making dynamics can thus be interpreted as minimizing an effective cost function $\mathcal{C} \sim \sum_l (A\tau_l + B\Delta r_l)$, where $\tau_l$ and $\Delta r_l$ represent the waiting time and jump length during each transition.

\begin{figure}
	\includegraphics[width=16cm]{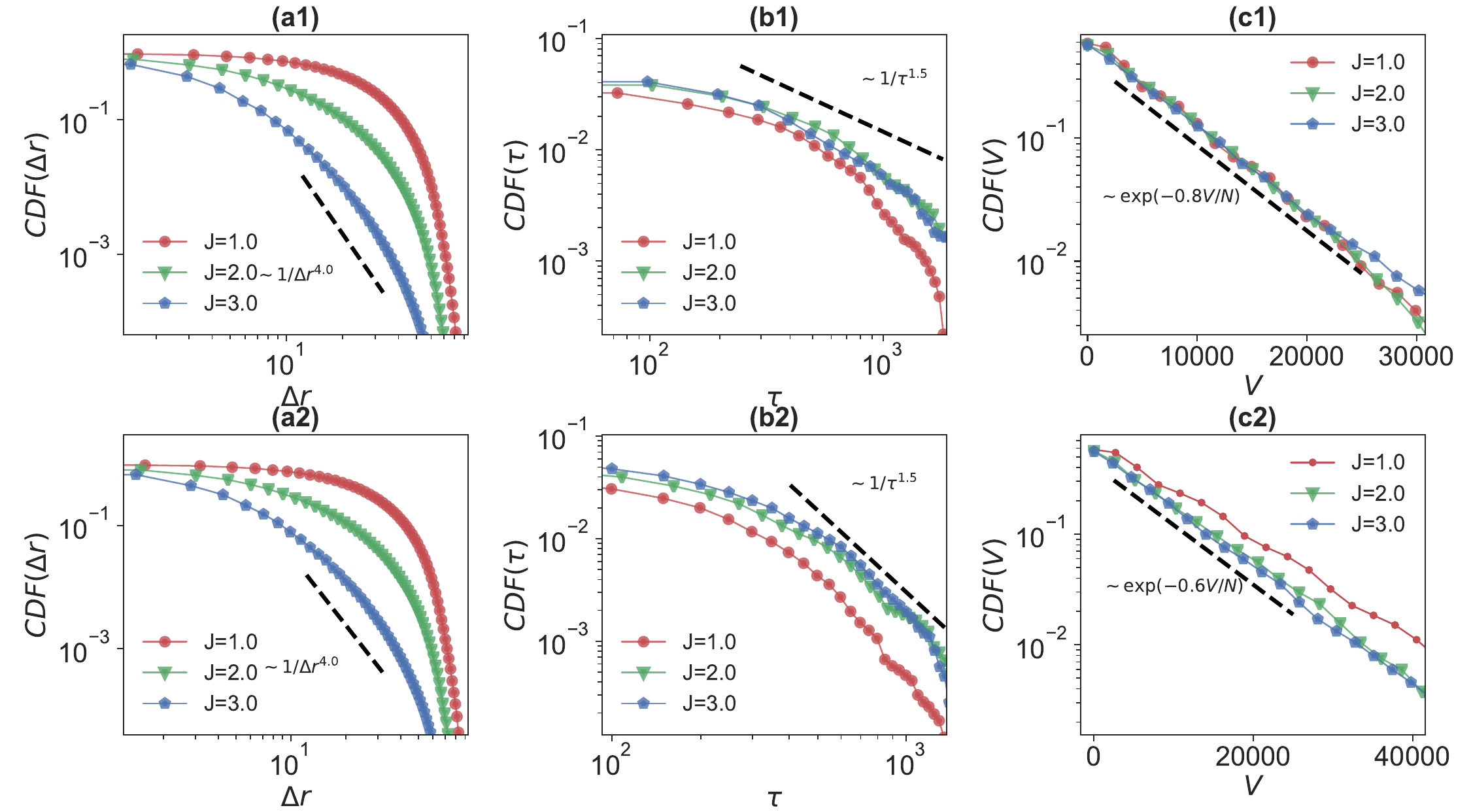} 
	\caption{Cumulative distribution functions for travel distance $\Delta r$, waiting time $\tau$ and visitation frequency $V$. All distribution functions are obtained after $T=10L$ number of steps with $L=70$ and $D=20$. Top panels: Results for the case in which agents remain in the queue to satisfy a given demand, regardless of their waiting time. Bottom panels: Results for the case in which agents, depending on their waiting time in the queue, either remain in the queue with probability $P_{wait}$ to satisfy a given demand or move to another location for this purpose. The reference lines with the mentioned exponents serve as a guide to the eye. 
	}\label{f5}
\end{figure}

In the following, we examine the results obtained from numerical simulations of the above model. Figure \ref{f3} shows the key quantities used to compare the dynamics of the model under variation of different parameters. One of the most important measures is the consumed energy $\langle E\rangle$, which is proportional to the total path length traveled in satisfying the demands. Other relevant quantities include the mean squared displacement from the origin $\langle r^2\rangle$ and the number of distinct visited sites $V$.

As observed, all these quantities exhibit power-law behavior within a certain range of the parameter $J$, which characterizes the relative importance of the destination–home distance $r_{b,O}$ compared to the origin–home distance $r_{a,O}$. It should be noted that, due to finite-size effects, the mean distance from the origin tends to saturate at a constant value. Similarly, the number of distinct visited sites is expected to approach a constant value at long times.

In Fig. \ref{f3} we compare the case in which the waiting time in the queue does not affect the dynamics ($\gamma = 0$) with the case where the waiting time is relevant ($\gamma = 1$). In the latter case, the agent may leave the queue by comparing it with a characteristic threshold $\tau_0$ and satisfy the same demand at another location. We see that the value of $\gamma$ does not change significantly the scaling exponents in this regime. For nonzero $\gamma$, the quantity most affected in terms of its power-law behavior is the mean-squared distance from the origin. 

Figure \ref{f4} shows the exploration probability as a function of the mean number of distinct visited sites. The tails exhibit power-law behavior, with exponents that vary with $J$. The same figure also presents efficiency metrics: (i) the ratio of the distance from the origin to the consumed energy $\eta_r$, and (ii) the ratio of the number of distinct visited sites to the consumed energy $\eta_v$, both evaluated over time for different values of the control parameter. These quantities exhibit behaviors that are similar to those observed in the single-particle dynamics. An optimal regime emerges within a specific range of the parameter $J\sim 2$.

The cumulative distribution functions of step length $\Delta r$, waiting times $\tau$, and the number of distinct visited sites $V$ for the two cases $\gamma = 0$ and $\gamma=1$ are shown in Fig. \ref{f5}. As can be seen, the dominant behavior for distributions of $\Delta r$ and $\tau$ is close to power-law, with exponents that depend on the control parameter $J$. For the number of visited sites $V$ we observe an exponential distribution with a decay rate that is smaller as expected for $\gamma=1$. Moreover, the qualitative behavior is not very sensitive to interactions that are induced by the waiting times for different values of $\gamma=0,1$.

\section{Conclusion}\label{S4}
We showed that key statistical features of human trajectories can be reproduced by a simple dynamical rule that depends on the ratio of the initial and final distances of a trip from the trajectory’s starting point (home). This hypothesis is supported by empirical data, which reveal a strong correlation between the two distances in a trip. We used a system of interacting agents that are to satisfy a list of demands and studied the impact of interactions and distributions of suppliers on the statistics of the trajectories. 

The above rule for accepting a trip can be readily incorporated into any movement model alongside other relevant dynamical mechanisms. For example, exploratory trips or returns to previously visited locations can be accepted with a probability that depends on the ratio of the endpoint distances from home (or a reference position). In the present work, we assumed that the available energy resources are unconstrained. It would be also interesting to investigate how imposing energetic constraints on mobility would modify the results. Because each trip carries a distance-proportional cost, the same rule offers a natural setting to study the energy efficiency of urban movement \cite{Biazzo2020,Biazzo2021,Mohsen2025}.

\acknowledgments
This work was performed using the ALICE compute resources provided by Leiden University.

\appendix

\section{Details of the empirical data}\label{app}
The GPS trajectory data used in this section were obtained from the GeoLife project by Microsoft Research Asia \cite{data-1,data-2,data-3}. The dataset was collected by 182 users between April 2007 and August 2012 and consists of 17,621 individual trajectories with a total travel distance of approximately $1.29\times10^{6}$ kilometers and a total duration of 50,176 hours. Each trajectory is represented as a sequence of GPS points recorded along the movement path, with each point containing latitude, longitude, and altitude information. 
The dataset captures a wide range of outdoor human mobility behaviors, including daily routines such as commuting between home and work, as well as entertainments and physical activities like shopping, sightseeing, dining, hiking, and cycling.
Although the trajectories are distributed across more than 30 cities in China and several locations in the United States and Europe, the majority of the data were collected in Beijing, China.

For this study, each trajectory was processed to obtain two basic measures: for each step $n$ which is defined by two successive GPS points $(a,b)$, we obtain the radial distances $r_{a,O}, r_{b,O}$ of points $a, b$ from the starting location $O$ of the trajectory. These quantities allow us to study how local movement steps change with distance from the origin along real human trajectories. The data and codes we used in this study are available at \cite{github-link}.


\begin{thebibliography}{prsty}



\bibitem{Gonzalez2008} M. C. Gonzalez, C. A. Hidalgo, and A. L. Barabási, ``Understanding individual human mobility patterns,'' Nature \textbf{453}, 779 (2008).

\bibitem{Batty2013} M. Batty, \textit{The New Science of Cities} (MIT Press, 2013).

\bibitem{Kraemer2020} M. U. Kraemer \textit{et al.}, ``The effect of human mobility and control measures on the covid-19 epidemic in china,'' Science \textbf{368}, 493 (2020).

\bibitem{Bettencourt2021} L. Bettencourt, \textit{Introduction to Urban Science: Evidence and Theory of Cities as Complex Systems} (MIT Press, 2021).

\bibitem{Gonzalez2022} D. Rybski and M. C. González, ``Cities as complex systems—Collection overview,'' PLoS ONE \textbf{17}, e0262964 (2022).

\bibitem{Barbosa2018} H. Barbosa \textit{et al.}, ``Human mobility: Models and applications,'' Phys. Rep. \textbf{734}, 1 (2018).

\bibitem{Pappalardo2023} L. Pappalardo, E. Manley, V. Sekara, and L. Alessandretti, ``Future directions in human mobility science,'' Nat. Comput. Sci. \textbf{3}, 588 (2023).

\bibitem{Barthelemy2016} M. Barthelemy, \textit{The Structure and Dynamics of Cities} (Cambridge University Press, 2016).

\bibitem{Chowell2003} G. Chowell, J. M. Hyman, S. Eubank, and C. Castillo-Chavez, ``Scaling laws for the movement of people between locations in a large city,'' Phys. Rev. E \textbf{68}, 066102 (2003).

\bibitem{Brockmann2006} D. Brockmann, L. Hufnagel, and T. Geisel, ``The scaling laws of human travel,'' Nature \textbf{439}, 462 (2006).

\bibitem{Alessandretti2017} L. Alessandretti, P. Sapiezynski, S. Lehmann, and A. Baronchelli, ``Multi-scale spatio-temporal analysis of human mobility,'' PLoS ONE \textbf{12}, e0171686 (2017).

\bibitem{Alessander2020} L. Alessandretti, U. Aslak, and S. Lehmann, ``The scales of human mobility,'' Nature \textbf{587}, 402 (2020).

\bibitem{Bettencourt2007} L. M. A. Bettencourt, J. Lobo, D. Helbing, C. Kühnert, and G. B. West, ``Growth, innovation, scaling, and the pace of life in cities,'' Proc. Natl. Acad. Sci. U.S.A. \textbf{104}, 7301 (2007).

\bibitem{Metzler2000} R. Metzler and J. Klafter, ``The random walk's guide to anomalous diffusion: a fractional dynamics approach,'' Phys. Rep. \textbf{339}, 1 (2000).

\bibitem{Rhee2011} I. Rhee, M. Shin, S. Hong, K. Lee, S. J. Kim, and S. Chong, ``On the Levy-walk nature of human mobility,'' IEEE/ACM Trans. Netw. \textbf{19}, 630 (2011).

\bibitem{Zhao2015} K. Zhao, M. Musolesi, P. Hui, W. Rao, and S. Tarkoma, ``Explaining the power-law distribution of human mobility through transportation modality decomposition,'' Sci. Rep. \textbf{5}, 9136 (2015).

\bibitem{Gallotti2016} R. Gallotti, A. Bazzani, S. Rambaldi, and M. Barthelemy, ``A stochastic model of randomly accelerated walkers for human mobility,'' Nat. Commun. \textbf{7}, 12600 (2016).

\bibitem{Song2010} C. Song, T. Koren, P. Wang, and A.-L. Barabási, ``Modelling the scaling properties of human mobility,'' Nat. Phys. \textbf{6}, 818 (2010).

\bibitem{Pappalardo2015} L. Pappalardo, F. Simini, S. Rinzivillo, D. Pedreschi, F. Giannotti, and A.-L. Barabási, ``Returners and explorers dichotomy in human mobility,'' Nat. Commun. \textbf{6}, 8166 (2015).

\bibitem{uvlaw} M. Schläpfer \textit{et al.}, ``The universal visitation law of human mobility,'' Nature \textbf{593}, 522 (2021).

\bibitem{switching} L. Zhong, L. Dong, Q. Wang, C. Song, and J. Gao, ``Switching exploration modes in human mobility,'' arXiv:2503.10969 (2025).

\bibitem{Wu2014} L. Wu, Y. Zhi, Z. Sui, and Y. Liu, ``Intra-urban human mobility and activity transition: Evidence from social media check-in data,'' PLoS ONE \textbf{9}, e97010 (2014).

\bibitem{Lee2015} M. Lee and P. Holme, ``Relating land use and human intra-city mobility,'' PLoS ONE \textbf{10}, e0140152 (2015).

\bibitem{Baumann2025} V. Baumann, J. Dambacher, M. F. L. Ruitenberg \textit{et al.}, ``Towards a characterization of human spatial exploration behavior,'' Behav. Res. \textbf{57}, 65 (2025).

\bibitem{Li2025} Q. Li \textit{et al.}, ``Slower searching yields higher efficiency: A case study of taxi drivers,'' Proc. Natl. Acad. Sci. U.S.A. \textbf{122}, e2502965122 (2025).

\bibitem{kolbl2003} R. Kölbl and D. Helbing, ``Energy laws in human travel behaviour,'' New J. Phys. \textbf{5}, 48 (2003).

\bibitem{kolbl2021} R. Kölbl and M. Kozek, ``A physiological model of human mobility: A global study,'' Humanit. Soc. Sci. Commun. \textbf{8}, 1 (2021).

\bibitem{Wang2022} W. Wang and T. Osaragi, ``Daily human mobility: A reproduction model and insights from the energy concept,'' ISPRS Int. J. Geo-Inf. \textbf{11}, 219 (2022).

\bibitem{inflation} L. Zhong, L. Dong, Q. Wang, C. Song, and J. Gao, ``Universal spatial inflation of human mobility,'' arXiv:2406.06889 (2024).

\bibitem{data-1} Y. Zheng, L. Zhang, X. Xie, and W.-Y. Ma, ``Mining interesting locations and travel sequences from GPS trajectories,'' in \textit{Proceedings of the 18th International Conference on World Wide Web (WWW 2009)} (ACM, New York, 2009), p. 791.

\bibitem{data-2} Y. Zheng, Q. Li, Y. Chen, X. Xie, and W.-Y. Ma, ``Understanding mobility based on GPS data,'' in \textit{Proceedings of the 10th International Conference on Ubiquitous Computing (UbiComp 2008)} (ACM, New York, 2008), p. 312.

\bibitem{data-3} Y. Zheng, X. Xie, and W.-Y. Ma, ``GeoLife: A collaborative social networking service among user, location and trajectory,'' IEEE Data Eng. Bull. \textbf{33}, 32 (2010).

\bibitem{Li2017} R. Li \textit{et al.}, ``Simple spatial scaling rules behind complex cities,'' Nat. Commun. \textbf{8}, 1841 (2017).


\bibitem{Biazzo2020} I. Biazzo and A. Ramezanpour, ``Efficiency and irreversibility of movements in a city,'' Sci. Rep. \textbf{10}, 1 (2020).

\bibitem{Biazzo2021} I. Biazzo, M. Ghasemi Nezhadhaghighi, and A. Ramezanpour, ``Entropy production of selfish drivers: implications for efficiency and predictability of movements in a city,'' J. Phys. Complex. \textbf{2}, 035026 (2021).

\bibitem{Mohsen2025} M. Ghasemi Nezhadhaghighi and A. Ramezanpour, ``Efficiency of energy-consuming random walkers: Variability in energy helps,'' Phys. Rev. E \textbf{111}, 014301 (2025).

\bibitem{Andreotti2025} E. Andreotti, U. Marquis, and R. Gallotti, ``Scale-free points-of-interest distribution in a city emerging from homogeneous Poissonian point processes,'' arXiv:2509.01699 (2025).

\bibitem{github-link} "https://github.com/aramezanpour/distance-from-home-matters/".


\end{thebibliography}
\end{document}